\documentstyle[12pt,epsfig]{article}
\def\beq{\begin{equation}}
\def\eeq{\end{equation}}
\def\be{\begin{equation}}
\def\ee{\end{equation}}
\def\bea{\begin{eqnarray}}
\def\eea{\end{eqnarray}}

\newcommand{\gsim}{\lower.7ex\hbox{$\;\stackrel{\textstyle>}{\sim}\;$}}
\newcommand{\lsim}{\lower.7ex\hbox{$\;\stackrel{\textstyle<}{\sim}\;$}}

\begin{document}

\begin{center}
{\LARGE\bf
Neutrino oscillations in low density medium} 

\vspace{0.6cm}
A. N. Ioannisian$^{a,b}$  and A. Yu. Smirnov$^{c,d}$\\ 
\vspace{0.3cm}
{\it $^a$Yerevan Physics Institute, Alikhanian Br.\ 2, 375036 Yerevan, Armenia}\\
{\it $^b$Institute for Theoretical Physics and Modeling, 375036 Yerevan, Armenia}\\
{\it $^c$ Department of Physics, Tokyo Metropolitan University,
Hachioji, Tokyo 192-0397, Japan}\\
{\it $^d$ ICTP, Strada Costiera 11, 34014 Trieste, Italy} \\
\end{center}
\begin{abstract}
For the case of small matter effects: $V \ll \Delta m^2/2E$, where $V$ is 
the matter potential,   
we develop the perturbation theory using $\epsilon \equiv 2VE/\Delta m^2$ 
as the expansion parameter. We derive simple and physically transparent 
formulas for the oscillation probabilities in the lowest order in $\epsilon$    
which are valid for  arbitrary density profile.  
The formulas  can be applied for propagation of the solar 
and supernova neutrinos in matter of the Earth,  
substantially simplifying numerical calculations.  
Using these formulas we study  sensitivity of the oscillation effects 
to structures of the density profile situated at  different distances 
from the detector $d$.  We show that 
for the mass-to-flavor state transitions, {\it e.g.}, $\nu_2 \to \nu_e$,  
the sensitivity is suppressed for remote structures:  
$d > l_{\nu} E/\Delta E$, where  $l_{\nu}$ is the oscillation length  
and $\Delta E/E$ is the energy resolution of detector.  

\end{abstract}


\section{Introduction}
\label{sec1}

For the LMA  oscillation parameters 
the oscillations of  solar and supernova (low energy) 
neutrinos inside the Earth 
occur in the weak matter effect  regime. 
That is, the matter potential, $V$, is much smaller than 
the kinetic energy of the neutrino system: 
\be
V \ll \frac{\Delta m^2}{2E}, 
\label{cond}
\ee
where 
\be 
V(x) = \sqrt{2} G_F N_e(x), 
\ee 
$G_F$ is the Fermi constant,  $N_e(x)$ is the number density of the electrons, 
and $\Delta m^2 = m^2_2 - m^2_1$ is the mass squared difference. 

In this case one  can introduce a small parameter 
\be 
\epsilon(x) = {2 E V(x) \over \Delta m^2 } \ll 1
\label{eps}
\ee
and develop the  perturbation theory in $\epsilon(x)$. 

Neutrino oscillations in the weak matter effect regime 
have been discussed extensively before \cite{w} - \cite{IS},   
in particular,  
for the solar and supernovae neutrinos propagating in the 
matter of the Earth. The previous work has been done mainly in the 
approximation of density profile which consists 
of several layers with constant density. 

Here we derive general formula which is valid for  
arbitrary density profile provided that the condition (\ref{cond}) 
is satisfied.  The formula is the generalization of our result 
in~\cite{IS} obtained for thin layers of matter.

\section{ $\epsilon$- perturbation theory}

We will consider two active neutrino mixing
\bea
\nu_f = U(\theta) \nu_{mass}, 
\eea 
where $\nu_f \equiv (\nu_e, \nu_a)^T$, $\nu_{mass} \equiv (\nu_1, \nu_2)^T$   
are the flavor and mass states correspondingly,   and  
\begin{equation} 
U \equiv \left( \begin{tabular}{cc} 
$\cos \theta$ & $\sin \theta$ \\
$ - \sin \theta$ & $\cos \theta$     
\end{tabular}
\right) 
\label{matrix1}
\end{equation}
is the mixing matrix in vacuum. 
In general $\nu_a$ is a combination of $\nu_{\mu}$ and $\nu_{\tau}$.

We will first find the evolution matrix for the mass eigenstates and 
then make projection onto the flavor states. 
Evolution of the mass states is given by the equation 
\begin{equation}
i \frac{d \nu_{mass}}{dx} = H(x) \nu_{mass}  
\label{evoleq}
\end{equation}
with the Hamiltonian 
\begin{equation}
\label{H}
H(x)  =   
\left( \! \! \begin{tabular}{cc} 
0 & 0 \\
0 & $ {\Delta m^2 \over 2 E} $   
\end{tabular}  \! \! \right)  + 
U^\dagger \left(  \! \! \begin{tabular}{cc} 
$V(x)$ & 0 \\
0 & 0   
\end{tabular}  \! \! \right) U. 
\end{equation}
The Hamiltonian  can  be rewritten as 
\begin{equation}
\label{Hdiag}
H(x) =    U^\prime(x) \left(  \! \! \begin{tabular}{cc} 
0 & 0 \\
0 & $ \Delta_m(x) $   
\end{tabular}  \! \! \right) U^{\prime \dagger}(x) , 
\end{equation}
where 
\be
\Delta_m(x) \equiv \frac{\Delta m^2}{2E} \sqrt{(\cos 2 \theta - 
\epsilon(x))^2+\sin^2 2 \theta}\ \ \ \ 
\label{split}
\ee
is  the difference of energies of the neutrino eigenstates in 
matter $\nu_m \equiv (\nu_{1m}, \nu_{2m})^T$,      
and $U^\prime \equiv  U( 
\theta^\prime)$ 
is the  instantaneous mixing matrix of the mass states in matter with 
the angle $\theta^{\prime}(x)$ determined by 
\be
\sin 2 \theta^\prime(x) = {\epsilon(x) \ \sin 2 \theta \over 
\sqrt{(\cos 2 \theta - \epsilon(x))^2 + \sin^2 2\theta}}. 
\label{angle}
\ee
Using (\ref{split}) the expression for the mixing angle 
of the mass states in matter,  $\theta^\prime(x)$,   
can be rewritten as 
\be
\sin 2 \theta^\prime  = \frac{V\sin  2\theta}{\Delta_m}   
= \epsilon \sin  2\theta_m.     
\label{angle}
\ee
In the last equality of  Eq. (\ref{angle}) 
$\theta_m$ is the mixing angle of the flavor states in matter. 
Taking into account relations  
$\nu_f = U(\theta_m) \nu_m$, and 
$\nu_{mass} = U(\theta^{\prime}) \nu_m$, 
we find that $\nu_f =U_m  (U^\prime)^{\dagger} \nu_{mass}$,  
and consequently, $U = U_m U^\prime$. Therefore  
\be
\theta^\prime =\theta_m - \theta. 
\ee 
The angle $\theta^{\prime}$ is small: $\sin 2 \theta^{\prime} \leq 
\epsilon$, as it should be in  the weak matter effect regime. \\

The  formal solution  of  the equation (\ref{evoleq}), that is,     
the evolution matrix from the initial 
point $x_0$ to the final point $x_f$,  can be written as 
\begin{equation}
\label{chronology}
S(x_0 \to \ x_f)  = T e^{- i \int_{x_0}^{x_f} H(x) \  dx},  
\end{equation}
where $T$ means the chronological ordering. 
Let us divide a trajectory of  neutrinos  
into $n$ equal parts (layers) of the size, $\Delta x$, so that 
$n = (x_f - x_0)/ \Delta x$, and assume 
constant density inside each layer. 
Then for the evolution matrix we obtain 
\be
S(x_0 \to \ x_f)   = 
      e^{-i H(x_n) \ \Delta x} \cdot e^{-i H(x_{n-1}) \ \Delta x} \cdots 
      e^{-i H(x_j) \ \Delta x} \cdots e^{-i H(x_1) \ \Delta x} \\
\label{layers1}
\ee
($x_n \equiv x_f$).  
According to (\ref{Hdiag}) it can be written as 
\be  
S(x_0 \to \ x_f) =   U^\prime_n D_n U_n^{\prime \dagger} \cdot  
      U^\prime_{n-1} D_{n-1} U_{n-1}^{\prime \dagger} 
      \cdots  U^\prime_j D_j U_j^{\prime \dagger} \cdots 
      U^\prime_1 D_1 U_1^{\prime \dagger},  
\label{Sprod}
\ee
where 
\begin{equation}
D_j \equiv \left( \! \begin{tabular}{cc} 
       1 & 0 \\
       0 & $  e^{i \phi^m_j}$ 
                        \end{tabular} 
                        \right), ~~~~~~\phi^m_{j} \equiv \Delta_m(V_j) 
\Delta x
\label{DD}
\end{equation}
is the evolution matrix of the matter eigenstates 
in $j$-th layer,   
$\phi^m_{j}$  
is the relative phase between the matter eigenstates acquired 
in the layer $j$, and $V_j$ is the value of  potential in the 
$j$-th  layer.

The mixing matrix $U^\prime_j $ is given by  
\be 
U^\prime_i = \left( \! \begin{tabular}{cc} 
       $\cos \theta^\prime_j$ & $\sin \theta^\prime_j$ \\
       $- \sin \theta^\prime_j$ & $\cos \theta^\prime_j$ 
                        \end{tabular}  \! \! \! \!
                        ~~\right) \ ,  
\label{matr-i}
\end{equation}
where $\theta_j$ is the mixing angle of the mass states in the layer $j$.

Structure of the expression (\ref{Sprod}) is rather transparent: 
it is the product of the blocks 
($U^\prime_j D_j U_j^{\prime \dagger}$)
for all the layers. In each layer we first project the mass states onto the 
matter eigenstates $\nu_m$, then evolve the 
eigenstates  and then project back to the mass states.  

Using (\ref{DD}) and (\ref{matr-i})  we find expression for the $j$-th 
block: 
\be
U^\prime_j D_j U_j^{\prime \dagger} = D_j +G_j ~, 
\label{block-j}
\ee
where 
\be
G_j = \left(e^{i \phi_j} - 1 \right) 
\left[ \frac{1}{2} \sin 2\theta^{\prime}_j  
\left( 
\begin{tabular}{cc}
0 & 1 \\     
1 & 0
\end{tabular} 
\right) 
+ \sin^2 \theta^{\prime}_j 
\left( 
\begin{tabular}{cc}
1 & 0 \\                               
0 & - 1
\end{tabular} 
\right) 
\right]. 
\label{G-matr}
\ee
Notice that $G_j = O(\epsilon)$,  furthermore the second term in 
(\ref{G-matr}), being proportional  to $\sin^2 \theta^{\prime}$,  is 
of the order $\epsilon^2$. 

Inserting expression (\ref{block-j}) into (\ref{Sprod}) we find 
\begin{equation}
\label{ups}
S(x_0 \to x_n) = ( D_n + G_n )( D_{n-1} + G_{n-1} ) 
                        \cdots ( D_1 + G_1 ), 
\end{equation}
and in the form of series  in powers of $G = O(\epsilon)$ it can be 
written as  
\begin{eqnarray}
\label{expup}
S_(x_0 \to x_f) & = & 
                     D_n D_{n-1} \cdots D_1 \nonumber \\
&+&                \sum_{j=1}^n D_n \cdots D_{i+1} G_j D_{j -1} 
                        \cdots D_1 \nonumber
\\
 & +&                 
                         \sum_{j=1}^n \sum_{k=1}^{j-1} 
                         D_n \cdots D_{j+1}G_j D_{j-1} 
                         \cdots D_{k+1} G_k D_{k-1} \cdots D_1 +  
                        \cdots 
\end{eqnarray}
The products of $D_j$ which appear in this formula equal   
\be
\Pi_{j = k ... n} D_j = \exp \left(i \sum_{j = k ... n} \phi_j^m\right) = 
\exp \left(i \sum_{j = k ... n} \Delta_m(V_j) \Delta x\right).
\ee

In the limit $n \to \infty$ and $\Delta x \to 0$ 
the sums are substituted by the integrals:  
$\sum \Delta x  \to \int dx$.  So that   
\be
\Pi_{j = k ... n} D_j  \to 
\exp (i \phi^m_{x_k \to x_n} ), 
\ee
where 
\be 
\phi^m_{x_k \to x_n} \equiv  \int_{x_k}^{x_n} d x \Delta_m (x).
\label{phim}
\ee
Furthermore, 
\be
G_j \to i \Upsilon (x) dx, 
\ee
where according to (\ref{G-matr})
\be   
\Upsilon = \frac{1}{2} \sin 2\theta ~V(x)
\left(
\begin{tabular}{cc}
0 & 1 \\
1 & 0  
\end{tabular} 
\right)
+ \Delta_m(x) \sin^2 \theta^{\prime}(x) 
\left(
\begin{tabular}{cc}
1 & 0 \\
0 & -1
\end{tabular}   
\right).
\label{U-matr}
\ee
Here we have taken into account the relation (\ref{angle}). 

Substituting the sums by the integrals in Eq. (\ref{expup})  
we find the S-matrix in terms of $\Upsilon$: 
\begin{eqnarray}
\label{int}
S(x_0 \to x_f)
&=&
                        \left( \! \begin{tabular}{cc} 
       1 & 0 \\
       0 & $  e^{i \phi^m_{x_0 \to x_f}}$ 
                        \end{tabular} \! \! \! \!
                        \right) + i \int_{x_0}^{x_f} dx \
                        \left( \! \begin{tabular}{cc} 
       1 & 0 \\
       0 & $  e^{i \phi^m_{x \to x_f}}$ 
                        \end{tabular}  \! \! \! \!
                        \right)
                        \Upsilon(x)
                        \left( \! \begin{tabular}{cc} 
       1 & 0 \\
       0 & $  e^{i \phi^m_{x_0 \to x}}$ 
                        \end{tabular} \! \! \! \!
                        \right)  \nonumber
 \\                
 &-&
                         \int_{x_0}^{x_f} dx \int_{x_0}^x dy \ 
                         \left( \! \begin{tabular}{cc} 
       1 & 0 \\
       0 & $  e^{i \phi^m_{x \to x_f}}$ 
                        \end{tabular}  \! \! \! \!
                        \right)
                        \Upsilon(x)
                        \left( \! \begin{tabular}{cc} 
       1 & 0 \\
       0 & $  e^{i \phi^m_{y \to x}}$ 
                        \end{tabular} \! \! \! \!
                        \right)
                        \Upsilon(y)
                        \left( \! \begin{tabular}{cc} 
       1 & 0 \\
       0 & $  e^{i \phi^m_{x_0 \to y}}$ 
                        \end{tabular} \! \! \! \!
                        \right) + \cdots
\end{eqnarray}
Essentially we perform an  expansion of the $S$-matrix in powers of 
$\Upsilon$.

Let us take the zero and the first order terms in 
$\Upsilon$ in (\ref{int}) which correspond to 
the zero and first order terms in  $\epsilon$. 
We find 
\be
\label{A}
S(x_0 \to x_f) =   
\left( \! \begin{tabular}{cc} 
       1 & 0 \\
       0 & $  e^{i \phi^m_{x_0 \to x_f}}$ 
                        \end{tabular} \! \! \! \!
                        \right) 
        +  
                        i {\sin 2\theta \over 2} \int_{x_0}^{x_f} dx  V(x) 
                        \left( \! \! \begin{tabular}{cc} 
                               0 & $e^{i \phi^m_{x_0 \to x}}$ \\
                               $e^{i \phi^m_{x \to x_f}}$ &   0   
                        \end{tabular} \! \! \! \! \right).  
\nonumber 
\ee

An interesting feature of the formula  (\ref{A}) which leads to important 
consequences (see sec. 4) is that in the second term the 12-element 
is determined by the phase  acquired from the initial point 
$x_0$ to a given 
point $x$, whereas the 12-element  depends on the phase from a given 
point $x$ to the final point of evolution $x_f$.  

The key  feature of the method  is that we calculate the adiabatic 
phase  exactly, whereas the 
mixing and the amplitude of oscillations  
are found  approximately using the perturbation in $\epsilon$. 

Notice that the second term in $\Upsilon$ (\ref{U-matr}) (of the order 
$\epsilon^2$) is proportional to the diagonal matrix.  
The 11-element of this term  does not contain oscillatory factor and 
therefore,  being inserted in (\ref{int}), 
produces  the contribution to the $S$- matrix 
which is proportional to the length of the trajectory.    
It can be shown  that this large contribution is canceled precisely 
by the contribution from the term of the second order in 
$\Upsilon$ (second line in Eq. (\ref{int})).

\section{Probabilities of neutrino conversion}

Using the evolution matrix in the mass states basis (\ref{A}) we can 
calculate the amplitudes  and probabilities of  different transitions. 
The evolution matrix in the flavor basis, $S_f$,  can be obtained 
immediately: 
\be
S_f = U S U^{\dagger}. 
\ee
The evolution matrix from the mass states to  the flavor states, $S_{fm}$, 
is then 
\be
S_{fm} = U S,  
\ee
and $U$ is the vacuum mixing matrix (\ref{matrix1}).

Let us consider the most important examples.  
The amplitude of the mass-to-flavor transition, $\nu_i \to \nu_\alpha$, 
on the way from $x_0$ to $x_f$ is given by 
\be
A_{\nu_i \to \nu_\alpha} = U_{\alpha j } S_{ji}. 
\ee
Inserting the matrix (\ref{A}) in this expression we find
the amplitude of the $\nu_1 \to \nu_e$ oscillations in  the first 
order in $V$: 
\begin{equation}
A_{\nu_1 \to \nu_e} = 
                        \cos \theta + {i \over 2} \sin 2 \theta \ \sin \theta 
                        \int_{x_0}^{x_f} dx \ V(x) e^{i \phi^m_{x \to x_f} 
}. 
\label{ampl}
\end{equation}
Then the probability of the  $\nu_1 \to \nu_e$ oscillations equals
\be
\label{p1}
P_{\nu_1 \to \nu_e} =        
      \ \cos^2 \theta - {1 \over 2} \sin^2 2 \theta 
      \int_{x_0}^{x_f} dx \ V(x) \sin \phi^m_{x \to x_f} \ .
\ee

The probability of $\nu_2 \to \nu_e$ transition  
relevant for the solar neutrino oscillations 
in the Earth can be obtained immediately from the unitarity condition: 
\be
\label{p2}
P_{\nu_2 \to \nu_e} =        
      \ \sin^2 \theta + {1 \over 2} \sin^2 2 \theta 
      \int_{x_0}^{x_f} dx \ V(x) \sin \phi^m_{x \to x_f}. 
\ee
Then the regeneration parameter 
defined as $f_{reg} \equiv P_{\nu_2 \to \nu_e} -\sin^2 \theta$ 
(see, {\it e.g.}, \cite{GCS})  
equals 
\be
\label{reg}
f_{reg} = {1 \over 2} \sin^2 2 \theta 
\int_{x_0}^{x_f} dx \ V(x) \sin \phi^m_{x \to x_f} . 
\ee
Using expression for the phase  $\phi^m_{x \to x_f}$  
(\ref{phim}) we obtain explicitly  
\be
\label{reg1}
f_{reg} = {1 \over 2} \sin^2 2 \theta
\int_{x_0}^{x_f} dx \ V(x) \sin \int_{x}^{x_f} dy 
\frac{\Delta m^2}{2E} \sqrt{ \left(\cos 2 \theta -
\frac{2V(y) E}{\Delta m^2} \right)^2+\sin^2 2 \theta} . 
\ee

Let us underline that these expressions are valid for 
arbitrary density profile provided that the condition 
(\ref{cond}) is fulfilled. 

The terms which contain $V$ are implicitly of the order 
$\epsilon$. Indeed, noting that the phase $\phi_{x \to x_f}$ is 
proportional to $\Delta m^2/2E$ and  
changing  the integration variable   
$x \rightarrow x \Delta m^2/2E$, one obtains the factor  
$\epsilon(x)$. 
 
It is straightforward to show that the obtained formulas reproduce the well 
know results for particular density distributions.  
Thus,  for one layer with constant 
potential the Eq. (\ref{reg}) gives immediately:  
\bea
f_{reg}= \epsilon \sin^2 2\theta 
\sin^2\frac{ \pi L}{l_m} \ , 
\label{regold}
\eea
where  $l_m = 2\pi/ \Delta_m$ 
is the oscillation length in matter. 
For the  profile with $n$ symmetric shells 
we obtain  performing explicit integration in  (\ref{reg})
\be
f_{reg} = \frac{2 E}{\Delta m^2}\sin^2 2\theta \sin\Phi_0 
\sum^{n-1}_{i=0} \Delta V_i \sin\Phi_i \ , 
\label{regnewb}
\ee
where $\Phi_0$ is the total phase acquired along the trajectory 
and $\Phi_i$ is the phase acquired inside external radius of a given shell, 
$\Delta V_i$ is the jump of potential at border of the $i$-th 
shell~\cite{HLS}.  
In general, the formulas can be obtained from the adiabatic 
perturbation theory, performing integration by parts 
in Eq. (18) of ~\cite{HLS}. In the lowest order in $\epsilon$ 
the adiabatic perturbation theory and the $\epsilon$-perturbation 
theory coincide.

\section{Sensitivity of oscillations to structures of the density  
profile}

The formulas obtained in section 3 can be used to estimate  
a sensitivity of the oscillation effects to various  structures of  
density profile. 

Suppose that there is some structure in the density profile (or $V$) in 
the 
point $x$. Then according to (\ref{p1}), (\ref{p2}) or (\ref{reg}) this 
feature will be 
integrated with $\sin \phi_{x \to x_f}$ - the periodic function
whose phase  is acquired  from the   structure position to the 
detector: $d \equiv x_f - x$. The larger the distance and therefore the 
phase, the stronger averaging effect is expected. 
So that the effect of remote structures of the profile on the 
mass-to-flavor oscillation probabilities is  suppressed. 
 
To quantify this sensitivity  
let us  introduce the  energy resolution function 
$f(E^\prime, E)$ and perform averaging of the probability 
folded with $f$: 
\begin{equation}
{\overline P_{\nu_1 \to \nu_e}} = \int dE^\prime f(E^\prime, E) \ P_{\nu_1 \to
\nu_e}.  
\label{avv}   
\end{equation}
It is convenient to parameterize the effect of integration 
introducing the {\it attenuation} factor $F(d)$ in the probability  as   
\be
{\overline P_{\nu_1 \to \nu_e}} 
= \ \cos^2 \theta - {1 \over 2} \sin^2 2 \theta 
\int_{x_0}^{x_f} dx  V(x)   F(x_f - x)  \sin \phi_{x \to x_f} ,  
\label{avvf}
\ee
\noindent
so that in the absence of averaging  $F = 1$.

\begin{figure}[t]
\centerline{\epsfxsize=99mm\epsfbox{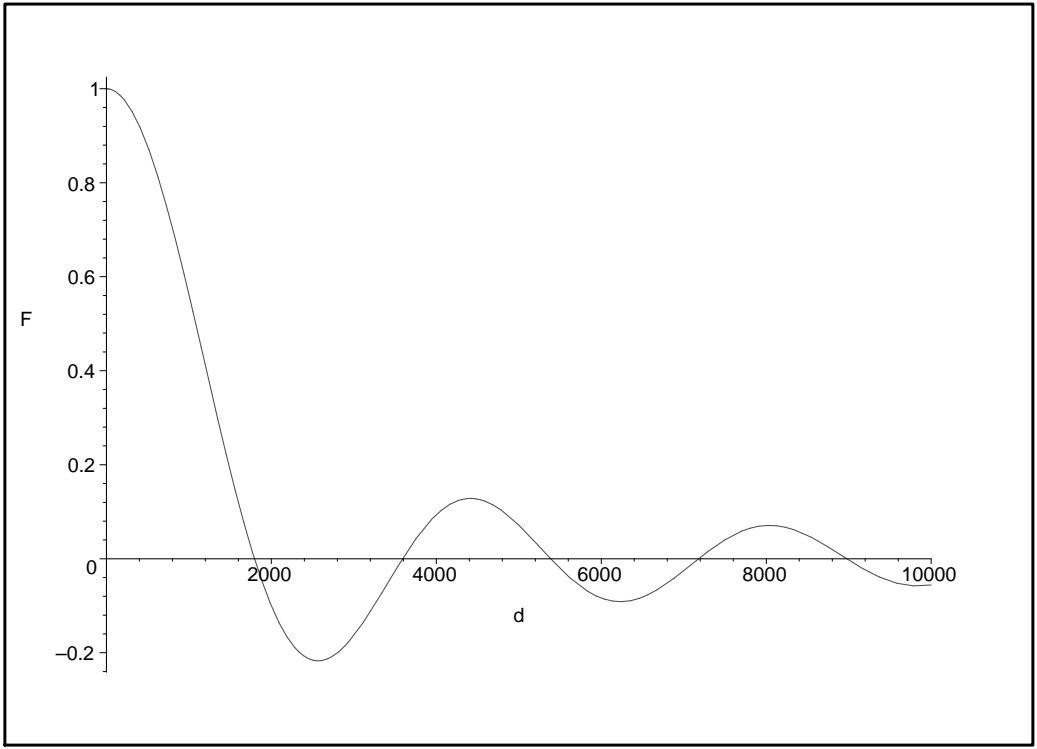}}
\caption{\small 
The attenuation factor $F$ as function of $d \equiv x_f-x $ for   
$E = 10$ MeV, $\Delta E = 2$ MeV, and 
$\Delta m^2 = 7 \cdot 10^{-5} $ eV$^2$.}
\label{fig1}
\end{figure}
Let us take  for simplicity  the box 
like resolution function $f (E^\prime, E)$  
for which   
\begin{equation}
{\overline P_{\nu_1 \to \nu_e}} = 
                        {1 \over \Delta E} \int_{E-{\Delta E \over 2}}^{E+
       {\Delta E \over 2}} dE P_{\nu_1 \to \nu_e} .
\end{equation}
Then assuming  that $ \Delta E \ll E$ and making use of 
the approximation $\Delta_m \simeq \Delta m^2[1- 
\epsilon \cos2\theta]/2E$  
we find 
\be
F(d)   =   
\frac{1}{Q(d)} \sin Q(d), ~~~~~
Q(d) \equiv { \pi d \Delta E \over l_{\nu} E } \ ,
\label{ffac}
\ee
where $l_{\nu} \approx l_m$ is the oscillation length (see Fig. 1). 

As follows from the  Fig. 1,  the factor $F(d)$ is  decreasing  
function of the distance from detector.   
The decrease of $F$ means that  contributions from the large 
distances to the integral (\ref{avvf}) are suppressed. 
According to Fig.~1 which corresponds to $\Delta E/ E = 0.2$, 
the effect of structures  at   distances 
above $1500$ km are attenuated by $F$ at least by factor 5 
in comparison with those situated  near the detector.
Correspondingly, the sensitivity to remote structures is much weaker.

The larger $\Delta E$, the smaller the width of the first peak 
of $F(d)$. As follows from (\ref{ffac}), the 
width and therefore the region of unsuppressed contributions 
due averaging are given by 
\be
d < l_{\nu}\frac{E}{\Delta E}. 
\ee

This attenuation effect allows us to  explain some features of the 
zenith angle distribution of the regeneration factor. No significant 
enhancement of the regeneration has been found in numerical calculations  
for the trajectories which cross the core of the Earth 
in comparison with  the trajectories crossing the mantle only,   
in spite of the fact that  density of the core,  
and therefore $\epsilon$,  are 2 - 3 times larger 
(see {\it e.g.} \cite{GCS}, \cite{cf}, \cite{sk-dn}).   
The explanation is straightforward: 
the border of the core is at about $D = 3000$ from the surface of 
the Earth. Therefore, according to Fig. 1 the effect of the structures 
from  such a distance is suppressed by the factor 3 - 5 which 
compensate the increase of $\epsilon$. On the other hand, 
the effect of even small  structures near the surface of the Earth can be   
substantial. 

The  conclusions we made so far  are valid for the 
oscillations of the mass states to the flavor states. 
The situation is different for the inverse -  flavor-to-mass states    
transitions. 
Indeed, for $\nu_e \to \nu_1$  the probability 
$P_{\nu_e \to \nu_1}$ has similar expression as in 
(\ref{p1}) but with phase acquired on the way from the initial 
point $x_0$ to the position of the structure $x$: $\phi_{x_0 \to x}$. 
Therefore with the flavor-to-mass transition probability one would 
probe structures at the opposite (to the detector) 
side of the Earth. 
This general consideration is in agreement with  our results 
for thin layers of matter ~\cite{IS}.

For the flavor to flavor transition we obtain the probability 
\be
\label{peef}   
P_{\nu_e \to \nu_e} = 1 - \sin^2 2 \theta 
\sin^2 \frac{\phi^m_{x_0 \to x_f}}{2}     
 - {1 \over 2} \sin^2 2 \theta \cos 2\theta 
\int_{x_0}^{x_f} dx \ V(x) [ \sin \phi^m_{x_0 \to x} +  
\sin \phi^m_{x \to x_f}] 
\ee
which is sensitive to structures near the surface from the 
both sides of the Earth. 

The detailed analysis of the averaging effects and other applications of 
the obtained results will be given elsewhere \cite{IS3}.

\section{Conclusion}

We have derived expressions for the oscillation probabilities in matter 
with  {\it arbitrary} density profile in the  weak matter effect regime:  
$V \ll \Delta m^2/ 2E$.  
An accuracy of these  expressions (\ref{p1}, \ref{p2}, \ref{reg}) is 
determined by the parameter $\epsilon$.

The results can be applied for the solar and supernova (low energy) neutrinos 
crossing the Earth.  They substantially simplify  numerical 
calculations of the oscillation effects.   

The obtained formulas reproduce the known probabilities for particular 
density distributions (one layer with constant density, 
several layers, {\it etc.}). 
 
The formulas have very simple structure and  can be used  efficiently 
for analysis of various effects. 
In particular, the sensitivity of the oscillation 
probabilities to various structures of the density profile 
can be easily evaluated. We have found that the 
mass-to-flavor transition probabilities  are sensitive to 
structures situated close enough to  detector.  
Effect of the remote structures is attenuated. 
The distance which can be viewed  by a detector is determined  
by the oscillation length divided by the energy resolution 
of the detector: $l_{\nu}E/\Delta E$.   These results can be used in 
future in the oscillation tomography of the Earth.

\section*{Acknowledgment} 

The work of A.N.I. was partially supported by SCOPES grant 7AMPJ062161.


\end{document}